\documentclass [preprint,showpacs,nofootinbib,superscriptaddress]{revtex4}
\usepackage{verbatim}
\usepackage{slashed}
\usepackage{epsfig}
\usepackage{graphicx}
\usepackage{amsmath}
\usepackage{mathrsfs}
\usepackage{bm}

\begin{document}


\title{Inclusive $J/\psi$ Production In $\Upsilon$ Decay  Via Color-singlet Mechanism}


\author{Zhi-Guo He}
\affiliation{Institute of High Energy Physics, Chinese Academy of
Science, P.O. Box 918(4), Beijing, 100049, China.\\
Theoretical Physics Center for Science Facilities,(CAS) Beijing,
100049, China.}

\affiliation{\small{\it{Departament d'Estructura i Constituents de
la Mat\`eria
                   and Institut de Ci\`encies del Cosmos}}\\
        \small{\it{Universitat de Barcelona}}\\
        \small{\it{Diagonal, 647, E-08028 Barcelona, Catalonia, Spain.}}\footnote{Present address}}

\author{Jian-Xiong Wang}

\affiliation{Institute of High Energy Physics, Chinese Academy of
Science, P.O. Box 918(4), Beijing, 100049, China.\\
Theoretical Physics Center for Science Facilities,(CAS) Beijing, 100049, China.}

\date{\today}

\begin{abstract}\vspace{5mm}
We reconsider the tree level color-singlet contribution for the
inclusive $J/\psi$ production in $\Upsilon$ decay with the
$\alpha_{s}^{5}$ order QCD process $\Upsilon\to J/\psi+c\bar{c}+g$
and $\alpha^{2}\alpha_s^{2}$ order QED processes
$\Upsilon\to\gamma^{\ast}\to J/\psi+c\bar{c}$ and $\Upsilon\to
J/\psi+gg$. It is found that the contribution of the
QED process is compatible with that of the QCD process,
and the numerical results for the QCD process alone is an order of magnitude smaller
than the previous theoretical predictions,
and our theoretical prediction in total is about an order of magnitude smaller than
the recent CLEO measurement on the branching fraction $\mathcal{B}(\Upsilon\to J/\psi+X)$.
It indicates that the $J/\psi$ production mechanism in $\Upsilon$ decay is not well
understood, and further theoretical work and experimental analysis are still necessary.
\end{abstract}


\maketitle
\section{Introduction}
Since the discovery of $c\bar{c}$ state $J/\psi$ and $b\bar{b}$
state $\Upsilon$ more than three decades ago, heavy-quarkonium
system has served as a good laboratory for testing QCD from both
perturbative and non-perturbative aspects. With the accumulation of
new experimental data and the development of interesting theory,
considerable attention has been attracted to study heavy-quarkonium
spectrum, decay and production (for a review see
\cite{Brambilla:2004wf}).

On the theoretical side, the non-relativistic QCD(NRQCD)\cite{Bodwin:1994jh}
effective field theory was introduced, based on which the production
and decay of heavy quarkonium can be calculated with a rigorous
factorization formalism. This formalism separates the physics on the
energy scale larger than the quark mass $m_Q$, related to the
annihilation or production of $Q\bar{Q}$ pair, from the physics on
the scale of $m_{Q}v^2$ order, relevant to the formation of the
bound state. Consequently, the inclusive production and decay rates
of heavy quarkonium are factorized into the product of
short-distance coefficients, which could be calculated
perturbatively as the expansion of $\alpha_{s}$, and the
corresponding long-distance matrix elements, which are determined by
some non-perturbative methods. The long-distance matrix elements are
weighted by the powers of $v$, the velocity of heavy quark in the
rest frame of the bound state. One important feature of NRQCD is
that it allows the contribution of $Q\bar{Q}$ pair in color-octet
configuration in short distance, and the color-octet state will
subsequently evaluate into physics state through the emission of
soft gluons.

The introduction of NRQCD has greatly improved our understanding of
the production mechanism of heavy quarkonium. One remarkable success
of NRQCD is that the transverse momentum ($p_t$) distributions of
$J/\psi$ and $\psi^{\prime}$ production at Fermilab
Tevatron\cite{Abe:1992ww} could be well described by the color-octet
mechanism\cite{Braaten:1994vv}. However, this mechanism could not
correctly explain the CDF measurements of $J/\psi$
polarization\cite{Affolder:2000nn}. Just about one or two years ago,
the next-to-leading order (NLO) QCD corrections to both the
color-singlet and color-octet processes have been obtained. For the
color-octet process\cite{Gong:2008ft}, it is found that the leading order (LO)
results are little changed when the NLO QCD corrections are taken into
account. In the color-singlet case, the theoretical predictions at
QCD NLO are significantly changed from the LO results on the $p_t$ distribution
and polarization of $J/\psi$\cite{Campbell:2007ws}. Although this still could not resolve
the puzzle. The large impact of the color-singlet NLO QCD
corrections on the LO results indicates that the contribution of
the color-octet mechanism may not as important as we expected
before. Furthermore, the theoretical predictions\cite{Artoisenet:2008fc} for
the $p_t$ distribution of $\Upsilon$ can compatible with the data of
$\Upsilon$ production at Tevatron\cite{Acosta:2001gv} within the
theoretical uncertainty when considering some of the important
next-to-next-to-leading-order (NNLO) $\alpha_{s}^{5}$ contribution. However, it still cannot explain the
recent polarization measurement by D0 Collaboration~\cite{:2008za}

In the case of $J/\psi$ production in $e^{+}e^{-}$ annihilation, the
existence of color-octet mechanism also faces to a challenge. The
NRQCD approach predicts that the $J/\psi$ production in $e^{+}e^{-}$
annihilation at LO in $\alpha_{s}$ is dominated by $e^{+}e^{-}\to
J/\psi+gg$, and $e^{+}e^{-}\to J/\psi+c\bar{c}$ and $e^{+}e^{-}\to
J/\psi+g$, in which the first two are color-singlet subprocesses and
the last one is color-octet subprocess. The color-octet
contribution\cite{Braaten:1995ez} predicts there is a peak in
$J/\psi$ momentum spectrum near the kinematic end point. Unfortunately,
The peak was not found in the experimental observation of
BABAR\cite{Aubert:2001pd} and Belle\cite{Abe:2001za}. By using the
soft-collinear effective theory (SCET), the color-octet
predictions\cite{Fleming:2003gt} could be softened, but it depends
on a unknown non-perturbative shape function. Belle also extended
their analysis by deriving associated $J/\psi$ production with
$c\bar{c}$ pair from inclusive $J/\psi$ production
production\cite{Abe:2002rb}. The NLO QCD calculations shown that
both $\sigma[e^{+}e^{-}\to J/\psi+c\bar{c}+X]$\cite{Zhang:2005cha,
Gong:2009ng} and $\sigma[e^{+}e^{-}\to
J/\psi+X_{\mathrm{non-}c\bar{c}}]$\cite{Ma:2008gq,Gong:2009kp} may
be explained by considering only the contribution of color-singlet
process. However, it point out in Ref.~\cite{Gong:2009ng} that the
color-octet contribution is still not yet completely ruled out due
to the incomplete measurement in the experimental analysis.

To improve our understanding of $J/\psi$ production mechanism, it
was proposed\cite{Cheung:1996mh,Napsuciale:1997bz} that the
$\Upsilon$ decay may provide an alternate probe of $J/\psi$
production in rich gluon environment. Experimentally, the branching
ratio of $\Upsilon \to J/\psi+X$ has already been reported to be
$(1.1\pm0.4\pm0.2)\times10^{-3}$ by CLEO based on about 20 events in
Ref.\cite{Fulton:1988ug}. The ARGUS Collaboration obtained an upper
limit of $0.68\times10^{-3}$\cite{Albrecht:1992ap} at $90\%$
confidence level. With about 35 times larger data sample than previous
work, an improved measurement of $J/\psi$ branching ratio and
momentum spectrum have been obtained recently by CLEO Collaboration
with $\mathcal{B}(\Upsilon\to
J/\psi+X)=(6.4\pm0.4\pm0.6)\times10^{-4}$\cite{Briere:2004ug}.
Theoretically, the color-octet prediction is
$\mathcal{B}(\Upsilon\to
J/\psi+X)\simeq6.2\times10^{-4}$\cite{Napsuciale:1997bz} with $10\%$
contribution from $\psi(2S)$ feed-down and another $10\%$ from
$\chi_{cJ}$\cite{Trottier:1993ze}. However, it was found that the
branching ratio of color-singlet process $\Upsilon\to
J/\psi+c\bar{c}g$ is about $5.9\times10^{-4}$\cite{Li:1999ar}, which
is also in agreement with experimental measurement. Although both the
color-singlet and color-octet decay modes may explain the total
decay rate independently, their predictions on the $J/\psi$ momentum
spectrum are significantly different. The maximum value of $J/\psi$
momentum in the color-singlet and color-octet process are 3.7 GeV
and 4.5 GeV respectively. The CLEO collaboration found that the experimental
result of $J/\psi$ momentum spectrum is much softer than color-octet
predictions and somewhat softer than color-singlet predictions. The
process $\Upsilon\to J/\psi+X$ also was studied in color evaporation
model\cite{Fritzsch:1978ey} more than thirty years ago, but this
model can not give systematic predictions of $J/\psi$ production.
Another early theoretical work on the process $\Upsilon\to J/\psi+X$
could be found in Ref.\cite{Bigi:1978tj}.

There is a very well agreement between the
LO color-singlet predictions\cite{Li:1999ar} and experimental
measurements\cite{Briere:2004ug}. But it seems difficult to understand the situation
in comparison with the case of the $J/\psi$ production
at B factories, where there are huge discrepancies between the LO theoretical predictions and
the experimental measurements. Therefore, we re-calculate the branching ratio of
color-singlet process $\Upsilon \to J/\psi+c\bar{c}+g$ in this paper. And the results show that
it is an order of magnitude smaller than the previous theoretical prediction ~\cite{Li:1999ar}.
Therefore, there is an order of magnitude discrepancy between the LO theoretical prediction
and experimental measurement for $\Upsilon\to J/\psi+X$ now.
To further clarify the situation, we also estimate the leading-order contribution of the QED
processes $\Upsilon\to\gamma^{\ast}\to J/\psi+c\bar{c}$
and $\Upsilon\to J/\psi+ gg$ at $\alpha^{2}\alpha_{s}^{2}$ order, in which the
process  $\Upsilon\to J/\psi+gg$ includes two gauge invariant subsets,
$\Upsilon\to\gamma^{\ast}\to J/\psi+gg$ and
$\Upsilon\to\gamma^{\ast}gg$ followed by $\gamma^{\ast}\to J/\psi$.
The final results show that the contribution from the QED processes are compatible
with that from the QCD process.

The rest of paper is organized as follows: In section II, the
basic formula and method used in the calculation are presented. In section III,
we describe the calculation on the branching ratio of the QCD process $\Upsilon\to
J/\psi+c\bar{c}+g$ and $J/\psi$ momentum spectrum. In section IV, we
estimate the contribution of the two QED processes $\Upsilon\to
J/\psi+c\bar{c}$ and $\Upsilon\to J/\psi+gg$. The final
results and summary are given in the last section.

\section{Description of Our basic calculation formula}

At leading order in $v_{Q}$, for S-wave heavy-quarkonium production
and decay, the color-singlet model predictions are equal to that
based on NRQCD effective theory. Then we express
$d\Gamma(\Upsilon\to J/\psi+X)$ as:
\begin{equation}
d\Gamma(\Upsilon\to
J/\psi+X)=d\hat{\Gamma}(b\bar{b}[^3S_1,\underline{1}]\to
c\bar{c}[^3S_1,\underline{1}]+X)
\langle\Upsilon|\mathcal{O}_1(^3S_1)|\Upsilon\rangle
\langle\mathcal{O}^{\psi}_1(^3S_1)\rangle,
\end{equation}
where $d\Gamma(b\bar{b}[^3S_1,\underline{1}]\to
c\bar{c}[^3S_1,\underline{1}]+X)$ represents color-singlet
$b\bar{b}$ pair in spin-triplet state decay into color-singlet
$c\bar{c}$ pair in spin-triplet state with anything, which is
calculated perturbatively, and
$\langle\Upsilon|\mathcal{O}_1(^3S_1)|\Upsilon\rangle$ and
$\langle\mathcal{O}^{\psi}_1(^3S_1)\rangle$ are the long-distance
matrix elements, which can be related to the nonrelativistic wave
functions as:
\begin{equation}
\langle\Upsilon|\mathcal{O}_1(^3S_1)|\Upsilon\rangle\simeq
\frac{3}{2\pi}|R_{\Upsilon}(0)|^{2},
\langle\mathcal{O}^{\psi}_1(^3S_1)\rangle =
\frac{9}{2\pi}|R_{\psi}(0)|^{2}.
\end{equation}

We employ spinor projection method\cite{Kuhn:1979bb} to calculate
the short-distance part $d\hat{\Gamma}$. In the nonrelativistic
limit, the amplitude of $b\bar{b}[^3S_1,\underline{1}]\to
c\bar{c}[^3S_1,\underline{1}]+X$ could be written
as\cite{Cho:1995vh}:
\begin{eqnarray}
&&\mathcal{M}(b\bar{b}[^3S_1,\underline{1}](p_0)\to
c\bar{c}[^{3}S,\underline{1}](p_{1})+X)=\sum_{s_1,s_2}
\sum_{i,l}\sum_{s_3,s_4}\sum_{k,l}\nonumber\\
&\times&\langle s_1;s_2\mid 1 S_z\rangle \langle 3i;\bar{3}j\mid
1\rangle\times\langle s_3;s_4\mid 1 S_z\rangle\langle 3k;\bar{3}l\mid 1\rangle\nonumber\\
 &\times& {\cal
M}(b_i(\frac{p_{0}}{2},s_1)\bar{b}_j(\frac{p_{0}}{2},s_2)\to
c_k(\frac{p_{1}}{2},s_3)\bar{c}_l(\frac{p_{1}}{2},s_4)+X)
\end{eqnarray}
where $\langle 3i;\bar{3}j\mid 1\rangle=\delta_{ij}/\sqrt{N_c}$,
$\langle 3k;\bar{3}l\mid 1\rangle=\delta_{kl}/\sqrt{N_c}$, $\langle
s_1;s_2\mid 1 S_z\rangle$, and  $\langle s_3;s_4\mid 1 S_{z}\rangle$
are the SU(3)-color, SU(2)-spin and angular momentum Clebsch-Gordan
(C-G) coefficients for $Q\bar{Q}$ projecting on certain appropriate
configurations at short distance. At leading order in $v_Q(Q=b,c)$,
the projection of spinors
$u(\frac{p_{0}}{2},s_1)\bar{v}(\frac{p_{0}}{2},s_2)$ and
$v(\frac{p_{1}}{2},s_3)\bar{u}(\frac{p_{1}}{2},s_4)$ could be
expressed as:
\begin{subequations}
\begin{equation}
\Pi_b=\sum_{s_1,s_2}\langle s_1;s_2\mid 1 S_z\rangle
u(\frac{p_{0}}{2},s_1)\bar{v}(\frac{p_{0}}{2},s_2)
=\frac{1}{2\sqrt{2}}\slashed{\epsilon}(S_z) (\slashed{p}_0-2m_b),
\end{equation}
\begin{equation}
\Pi_c=\sum_{s_1,s_2}\langle s_1;s_2\mid 1S_z\rangle
v(\frac{p_{1}}{2},s_3)\bar{u}(\frac{p_{1}}{2},s_4)
=\frac{1}{2\sqrt{2}}\slashed{\epsilon}(S_z) (\slashed{p}_1+2m_c),
\end{equation}
\end{subequations}
where $\epsilon(S_z)$ is the polarization vector of the heavy
quarkonium. For a spin=1 state with momentum $p$, the sum over its
all possible states $S_z$ is
\begin{equation}
\sum_{S_z}\epsilon_{\alpha}(S_z)\epsilon^{\ast}_{\beta}(S_z)=
(-g_{\alpha\beta}+\frac{p_\alpha p_\beta}{p^2})
\end{equation}

According to the spinor projection method, the
relation between $d\hat{\Gamma}$ and $|\mathcal{M}|^2$ for the
$b\bar{b}[^3S_1,\underline{1}]\to
c\bar{c}[^3S_1,\underline{1}]+X$  is
\begin{equation}
d\hat{\Gamma}(b\bar{b}[^3S_1,\underline{1}]\to
c\bar{c}[^3S_1,\underline{1}]+c\bar{c}+g)=\frac{1}{3}
\frac{1}{4m_b}\frac{\sum|\mathcal{M}|^2}{3m_bm_c(2N_c)^2}d\Phi_{n}
\end{equation}
where $\sum$ means to sum over all possible polarization states of
the particles in this process and $\Phi_{n}$ is the n-body phase
space. The factor $(1/2N_c)^2$ with $N_c=3$ comes from the
normalization factor of the NRQCD 4-Fermion operator.

Since our calculation gives different results from the previous theoretical prediction~\cite{Li:1999ar},
we further checked our results by using two different way to do all the calculations.
One is to apply the above formula to write a piece of program to do the calculations for each process described
in the following two section. Another is just using the Feynman Diagram Calculation (FDC)
Package~\cite{FDC} to generate all the needed Fortran source and then do the numerical calculation.
We obtained exactly the same results by using these two methods.
Moreover, to check gauge invariance, in the expression of FDC version source, the gluon polarization vector
is explicit kept and then is replaced by its 4-momentum in the final numerical calculation. Definitely
the result must be zero and our results confirm it.

\section{The QCD process $\Upsilon\to J/\psi+c\bar{c}+g$}
Now we proceed to calculate the total decay rate of $\Upsilon\to
J/\psi+c\bar{c}+g$ and its contribution to the $J/\psi$ momentum
spectrum. At leading order in $\alpha_{s}$, there are six Feynman
diagrams which are shown in Fig.~1. The amplitude $\mathcal{M}$
could be factorized as:
\begin{eqnarray}
\mathcal{M}(b\bar{b}[^3S_1,\underline{1}](p_{_0})\to
c\bar{c}[^3S_1,\underline{1}](p_{_1})+c(p_{_2})\bar{c}(p_{_3})+g(p_{_4}))
=\nonumber\\
\mathcal{M}_b(b\bar{b}[^3S_1,\underline{1}]\to
g^{\ast}g^{\ast}g) \times\mathcal{M}_c(g^{\ast}g^{\ast}\to
c\bar{c}[^3S_1,\underline{1}]+c\bar{c}),
\end{eqnarray}
in which the later one is universal for all the six diagrams and it
is
\begin{equation}
\mathcal{M}_c=\frac{g_s^2}{(p_2+p_1/2)^2(p_3+p_1/2)^2}
\bar{u}(p_2)\gamma^{\mu}\Pi_{c}\gamma^{\nu}v(p_3).
\end{equation}
The amplitude of $\mathcal{M}_b(b\bar{b}[^3S_1,\underline{1}]\to
g^{\ast}g^{\ast}g)$, for example for the first diagram, is
\begin{eqnarray}
\mathcal{M}^{1}_{b}=g_s^3C_{1}
\mathrm{Tr}[\Pi_b\gamma^{\mu}\frac{\frac{-\slashed{p}_0}{2}+\frac{\slashed{p}_1}{2}+\slashed{p}_3+m_b}
{(-p_0/2+p_1/2+p_3)^2-m_b^2}\gamma^{\nu}
\frac{\frac{\slashed{p}_0}{2}-\slashed{p}_4+m_b}
{(p_0/2-p_4)^2-m_b^2}\slashed{e}_3]
\end{eqnarray}
where $C_{1}$ is the corresponding color coefficient and $\slashed{e}_3$ is the polarization vector
of the real gluon. The amplitude
$\mathcal{M}^{i}_{b}$ for the other five diagrams could be obtained
in a similar way. An analytical expression of
$\sum|\mathcal{M}|^2$ is obtained in the calculation, but is too lengthy to be presented
here.

\begin{figure}
\begin{center}
\includegraphics[scale=0.5]{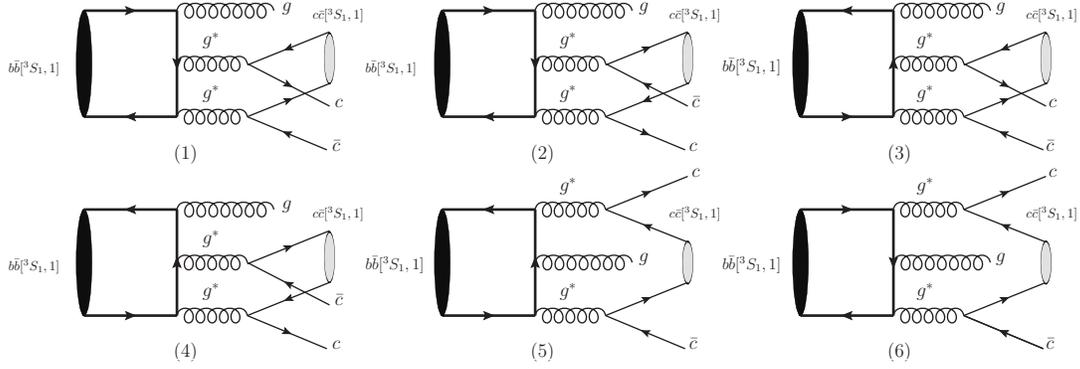}
\caption{The six Feynman diagrams for the short-distance process:
$b\bar{b}[^3S_1,1]\to c\bar{c}[^3S_1,1]+c\bar{c}+g$. }
\end{center}
\end{figure}

The four-body phase space $\Phi_{4}$ for
$b\bar{b}[^3S_1,\underline{1}]\to
c\bar{c}[^3S_1,\underline{1}]+c\bar{c}+g$ is defined as
\begin{equation}
d\Phi_{4}(p_{_0}\to
p_{_1}+p_{_2}+p_{_3}+p_{_4})=\prod_{k=1}^{4}\frac{d^3\vec{p}_{_k}}{(2\pi)^32E_k}(2\pi)^4
\delta^4(p_0-\sum_{k=1}^{4}p_{_k})
\end{equation}
There are many ways to perform the four-body phase-space
integration. Here we briefly introduce our methods. Using the two
following identical equation
\begin{equation}
\int\frac{d^4p_{_{234}}}{(2\pi)^4}(2\pi)^4\delta^{4}(p_{_{234}}-p_{_2}-p_{_3}-p_{_4})\equiv1,
\int\frac{d^4p_{_{34}}}{(2\pi)^4}(2\pi)^4\delta^{4}(p_{_{34}}-p_{_3}-p_{_4})\equiv1,
\end{equation}
we transform the four-body space into the combination of three
two-body phase spaces, which is given by
\begin{eqnarray}
&&d\Phi_{4}(p_{_0}\to
p_{_1}+p_{_2}+p_{_3}+p_{_4})=\frac{ds_{_{234}}}{2\pi}\frac{ds_{_{34}}}{2\pi}
\nonumber\\&&d\Phi_2(p_{_0}\to p_{_1}+p_{_{234}}) d\Phi_2(p_{_{234}}\to
p_{_2}+p_{_{34}})d\Phi_{2}(p_{_{34}}\to p_{_3}+p_{_4})
\end{eqnarray}
where $s_{_{234}}=p_{_{234}}^2,s_{_{34}}=p_{_{34}}^2$. The three
two-body phase spaces integration are described by the three-momenta
$\vec{p}_{_1},\vec{p}_{_2}^{\ast},\vec{p}_{_3}^{\ast\ast}$ and their
solid angle element
$d\Omega_{_0},d\Omega_{_{234}}^{\ast},d\Omega_{_{34}}^{\ast\ast}$ in
the rest frames of $p_{_0}$, $p_{_{234}}$, and $p_{_{34}}$
respectively. Then the expression of four-body phase space becomes
\begin{equation}
d\Phi_{4}=\int \frac{ds_{_{234}}}{2\pi} \int
\frac{|\vec{p}_{1}|}{8(2\pi)^2m_b}d\Omega_{_0}
\int\frac{ds_{_{34}}}{2\pi} \int
d\Omega_{_{234}}^{\ast}\frac{|\vec{p}_{_2}^{\ast}|}{4(2\pi)^2\sqrt{s_{_{234}}}}
\int\frac{|\vec{p}_{_3}^{\ast\ast}|}{4(2\pi)^2\sqrt{s_{_{34}}}}
d\Omega_{_{34}}^{\ast\ast}.
\end{equation}
where $|\vec{p}_{1}|$, $|\vec{p}_{_2}^{\ast}|$ and
$|\vec{p}_{_3}^{\ast\ast}|$ are given in the equations below
in the rest frame of $p_{0}$, $p_{_{234}}$ and $p_{_{34}}$ respectively
\begin{subequations}
\begin{equation}
|\vec{p}_{1}|=\frac{\sqrt{16m_b^4+(-4m_c^2+s_{_{234}})^2-8m_b^2(4m_c^2+s_{_{234}})}}
{4m_b}
\end{equation}
\begin{equation}
|\vec{p}_{_2}^{\ast}|=\frac{\sqrt{(s_{_{234}}-(m_c-\sqrt{s_{_{34}}})^2)
(s_{_{234}}-(m_c+\sqrt{s_{_{34}}})^2)}} {2\sqrt{s_{_{234}}}}
\end{equation}
\begin{equation}
|\vec{p}_{_3}^{\ast\ast}|=\frac{s_{_{34}}-m_c^2}{2\sqrt{s_{_{34}}}}
\end{equation}
\end{subequations}
The integration ranges of $s_{_{234}}$ and $s_{_{34}}$ are
\begin{equation}
4m_c^2<s_{_{234}}<(2m_b-2m_c)^2,m_c^2<s_{_{34}}<(\sqrt{s_{_{234}}}-m_c)^2.
\end{equation}

For space-symmetry, $d\Omega_{_0}$ and $d\phi_{_{234}}^{\ast}$ could
be integrated out directly then $|\mathcal{M}|^2$ only dependent on
five variables $s_{_{234}}$, $s_{_{34}}$, $\theta_{_{234}}^{\ast}$,
$\theta_{_{34}}^{\ast\ast}$, and $\phi_{_{34}}^{\ast\ast}$.
To get the total decay rate, the non-trivial integral with these
five variables is performed by three steps. First, we do the
integration $d\Omega_{_{34}}^{\ast\ast}$ in the rest frame of $p_{_{34}}$,
then we integrate out $s_{_{34}}$ and $\theta_{_{234}}^{\ast}$ in
the rest frame of $p_{_{234}}$, the last variable $s_{_{234}}$ is
integrated out in $\Upsilon$ rest frame. Since
$|\vec{p}_{1}|$ only depend on $s_{_{234}}$, the $J/\psi$ momentum spectrum
could be easily obtained by replacing $d s_{_{234}}$ with
$\frac{d s_{_{234}}}{d|\vec{p}_{1}|}d|\vec{p}_{1}|$.
The phase space integrations for the total rate and $J/\psi$
momentum spectrum are calculated numerically.

\begin{table}
\caption{The values of $f(r)$ for different $r={m_c}/{m_b}$} \small
\begin{tabular}{|c|c|c|c|c|c|c|c|}
\hline~ r ~&~ 0.275 ~&~ 0.296 ~&~ 0.317 ~&~ 0.327 ~&~ 0.338 ~&~ 0.361 ~&~ 0.381 ~ \\
\hline~ f(r)~&0.904&0.567&0.345&0.269&0.202&0.105&0.055\\
\hline
\end{tabular}
\end{table}

By dimension analysis, it is easy to represent the decay width
and differential decay width of $\Upsilon\to J/\psi+c\bar{c}+g$ as
\begin{subequations}
\begin{equation}
\Gamma(\Upsilon\to J/\psi+c\bar{c}+g)=\frac{\alpha_s^5}{m_b^{5}}f(r)
\frac{\langle\Upsilon|\mathcal{O}_1(^3S_1)|\Upsilon\rangle}{2N_c}
\frac{\langle\mathcal{O}^{\psi}_1(^3S_1)\rangle}{3\times2N_c}.
\end{equation}
\begin{equation}
\frac{d\Gamma}{d|\vec{p}_1|}(\Upsilon\to J/\psi+c\bar{c}+g)
=\frac{\alpha^s_5}{m_b^{6}}g(r,|\vec{p}_1|/m_b)
\frac{\langle\Upsilon|\mathcal{O}_1(^3S_1)|\Upsilon\rangle}{2N_c}
\frac{\langle\mathcal{O}^{\psi}_1(^3S_1)\rangle}{3\times2N_c}.
\end{equation}
\end{subequations}
where $r={m_c}/{m_b}$ and $f(r)$ are dimensionless, and $f(r)$ function
is same as $h(r)$ in Ref.\cite{Li:1999ar}. To
ensure the validity of our calculations, we use two different kinds
of computer codes for cross check and obtain exactly the same results for
$f(r)$ and $g(r,|\vec{p}_1|/m_b)$.
When $r=0.327$,
the decay width is
\begin{equation}
\Gamma(\Upsilon\to J/\psi+c\bar{c}+g)=
\frac{\alpha_{s}^5}{m_b^{5}}\frac{\langle\Upsilon|\mathcal{O}_1(^3S_1)|\Upsilon\rangle}{2N_c}
\frac{\langle\mathcal{O}^{\psi}_1(^3S_1)\rangle}{3\times2N_c}\times0.269.
\end{equation}
To compare our results with those in Ref.~\cite{Li:1999ar}, the
numerical results of $f(r)$ in the range of $0.275\leq r\leq0.381$
are listed in Tab.[I]. It is easy to see that the results of $f(r)$ are
about an order of magnitude smaller than that given in
Ref.\cite{Li:1999ar} and $f(r)$ changes a little sharper than that
when $r$ goes from 0.275 to 0.381.  Besides $f(r)$, the decay width
$\Gamma(\Upsilon\to J/\psi+c\bar{c}+g)$ is also dependent on the
choice of the values of the two long-distance matrix elements
$\langle\Upsilon|\mathcal{O}_1(^3S_1)|\Upsilon\rangle,
\langle\mathcal{O}^{\psi}_1(^3S_1)\rangle$, the coupling constant
$\alpha_{s}$ and the mass of b-quark. To reduce the uncertainty of
theoretical predictions, we normalize it by the decay width of
$\Upsilon\to \mathrm{light\; hadron}$, which includes two dominate
decay modes $\Upsilon\to ggg$ and $\Upsilon\to\gamma^{\ast}\to
q\bar{q}\;\mathrm(q=u,d,s,c)$. At leading order in $\alpha_{s}$ and
$v_b$, we have
\begin{subequations}
\begin{equation}
\Gamma(\Upsilon\to ggg)=
\frac{20\alpha_s^{3}(\pi^2-9)}{243m_b^2})\langle\Upsilon|\mathcal{O}_1(^3S_1)|\Upsilon\rangle,
\end{equation}
\begin{equation}
\Gamma(\Upsilon\to q\bar{q})= \frac{2\pi N_ce_q^2
e_b^2\alpha^2}{m_b^2}
\langle\Upsilon|\mathcal{O}_1(^3S_1)|\Upsilon\rangle.
\end{equation}
\end{subequations}
Then the normalized width $\Gamma^{c\bar{c}g}_{\mathrm{Nor}}$ is
given by
\begin{equation}
\Gamma^{c\bar{c}g}_{\mathrm{Nor}}=
\frac{f(r)\alpha_s^{5}\langle\mathcal{O}^{\psi}_1(^3S_1)\rangle}
{3(2N_c)^2(\frac{20}{243}\alpha_s^3(\pi^2-9)+\sum_{q}2\pi N_c e_q^2
e_b^2\alpha^2)m_b^3}
\end{equation}
and the branching ratio turns to be
\begin{equation}
\mathcal{B}(\Upsilon\to J/\psi+c\bar{c}+g)=
\Gamma^{c\bar{c}g}_{\mathrm{Nor}}\times\mathcal{B}(\Upsilon\to
\mathrm{light\;hadron}).
\end{equation}

Since the process $\Upsilon\to J/\psi+c\bar{c}+g$ can be viewed as
$\Upsilon\to g g^{\ast}g^{\ast}$ followed by $g^{\ast}g^{\ast}\to
J/\psi+c\bar{c}$, as suggested in Ref.\cite{Cheung:1996mh}, it is
reasonable to chose $\alpha_{s}(2m_c)=0.259$. Using
$e_u=\frac{2}{3}$,
$e_d=-\frac{1}{3}$,$e_s=-\frac{1}{3}$,$e_c=\frac{2}{3}$,
$e_b=\frac{1}{3}$, $\alpha=\frac{1}{128}$,
$r=\frac{1.548}{4.73}\simeq0.327$, $m_b=4.73\mathrm{GeV}$,
$|R_\psi(0)|^2=0.81\mathrm{GeV^{3}}$ being calculated in potential
model\cite{Eichten:1995ch} and
$\mathcal{B}(\Upsilon\to\mathrm{light\;hadron})=92\%$\cite{Amsler:2008zzb},
we predict
\begin{equation}
\mathcal{B}(\Upsilon\to J/\psi+c\bar{c}+g)=2.12\times10^{-5}
\end{equation}
The normalized $J/\psi$ momentum spectrum
$d\Gamma_{Nor}/{d|\vec{p}_1|}$ is shown in Fig.~3. It is easy
to see that the shape of the $J/\psi$ momentum spectrum is similar
with that in Ref.\cite{Li:1999ar}, although the prediction for the total
decay width is an order of magnitude smaller than the experimental data.

\section{The QED Process $\Upsilon\to J/\psi+X$ }
There are two QED processes $\Upsilon\to J/\psi+c\bar{c}$ and
$\Upsilon\to J/\psi+gg$ at the leading order in $\alpha_{s}$ and $\alpha$.
Both of them are considered in this work.
We will present a few simple steps and analytic results for them in the following.

\begin{figure}
\begin{center}
\includegraphics[scale=0.6]{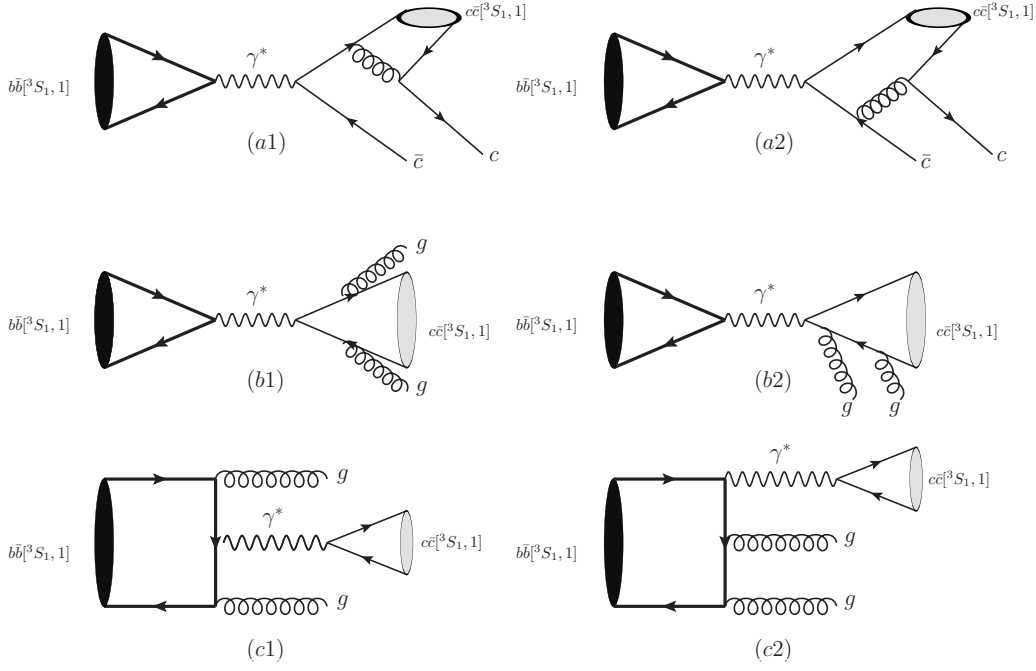}
\caption{The typical Feynman diagrams for the QED processes of
inclusive $J/\psi$ production: (a) $b\bar{b}[^3S_1,1]\to
\gamma^{\ast}\to\bar{c}[^3S_1,1]+c\bar{c}$, (b)
$b\bar{b}[^3S_1,1]\to \gamma^{\ast}\to\bar{c}[^3S_1,1]+gg$, (c)
$b\bar{b}[^3S_1,1]\to c\bar{c}[^3S_1,1]+gg$. }
\end{center}
\end{figure}

\subsection{$\Upsilon\to\gamma^{\ast}\to J/\psi+c\bar{c}$}
At the leading order, there are four Feynman diagrams for
$\Upsilon(p_0)\to\gamma^{\ast}\to J/\psi(p_1)+c(p_2)\bar{c}(p_3)$,
two of which are shown in Fig.~2a. The calculation procedure for this process
is very similar to  that for the $J/\psi$ production in association with
$c\bar{c}$ pair in $e^{+}e^{-}$ annihilation. The differential decay width is given by
\begin{eqnarray}
&&\frac{d\Gamma}{d|\vec{p}_{_1}|}(\Upsilon\to\gamma^{\ast}\to J/\psi+c\bar{c})
=\frac{2\pi C_AC_F^2e_b^2e_c^2\alpha^2\alpha_s^2
\langle\Upsilon|\mathcal{O}_1(^3S_1)|\Upsilon\rangle \langle\mathcal{O}^{\psi}_1(^3S_1)\rangle
\sqrt{x_1^2-4r^2}}
{9(2N_c)^2m_b^6 r\,{x_1}^4\,{( \kappa - x_1 ) }^3\, {( -2 + x_1 ) }^2\,{( \kappa + x_1 ) }^3}\nonumber\\&&
( 2\,\kappa x_1( -2{\kappa}^6( 1 + 2r^2 ) {x_1}^2 +
         {\kappa}^4( 6r^6( -4 + 3{x_1}^2 )  +
            2{x_1}^2( -4 + {x_1}^2( -2 + 9x_1 )  )
            -\nonumber\\&&
            4r^4( 16 + x_1( -16 + x_1( -8 + 9x_1 )  )  )  +
            r^2( -2 + x_1 ) ( 16 + x_1( -24 + x_1( -14 + 39x_1 )  )  )  )
            +\nonumber\\&&
         2{\kappa}^2{x_1}^2( 8{x_1}^2 + 7{x_1}^4 - 18{x_1}^5 -
            4r^6( -8 + x_1( 8 + x_1 )  )  +
            4r^4( 20 + x_1( -40 + x_1( 13 + 4x_1 )  )  )
            +\nonumber\\&&
            r^2( 32 + x_1( -96 + x_1( 60 + ( 76 - 37x_1 ) x_1 )  )  )  )  +
         {x_1}^4( 6r^6( 4 + {x_1}^2 )  +
            2{x_1}^2( -4 + {x_1}^2( -4 + 9x_1 )  )
            +\nonumber\\&&
            4r^4( 8 + x_1( 32 + ( -26 + x_1 ) x_1 )  )  +
            r^2( -32 + x_1( 128 + x_1( -124 - 60x_1 + 39{x_1}^2 )  )  )  )  )
            +\nonumber\\&&
      {( \kappa - x_1 ) }^3{( \kappa + x_1 ) }^3
       ( -6r^6( 4 + {x_1}^2 )  + 2{x_1}^2( 4 + {x_1}^2( -13 + 8x_1 )  )
4r^4( -16 + x_1+
         \nonumber\\&&( 16 + x_1( -4 + 5x_1 )  )  )  +
         r^2( -32 + x_1( 64 + x_1( 4 + ( 4 - 7x_1 ) x_1 )  )  )  )
       \log \frac{x_1-\kappa }{x_1+\kappa }  ),
\end{eqnarray}
where $C_A=3$ and $C_F=\frac{4}{3}$ are the color factors,
and there are $x_1=\sqrt{|\vec{p}_{_1}|^2+4m_c^2}/m_b$ and
$\kappa=\sqrt{(x_1+2r)(x_1-2r)(1+r^2-x_1)(1-x_1)}/{(1+r^2-x_1)}$.

Integrating $|\vec{p}_{_1}|$ numerically and normalizing
$\Gamma(\Upsilon\to\gamma^{\ast}\to J/\psi+c\bar{c})$ by
$\Gamma(\Upsilon\to\mathrm{light\; hadron})$, we obtain
\begin{equation}
\Gamma^{c\bar{c}}_{\mathrm{Normal}}=\frac{\Gamma(\Upsilon\to\gamma^{\ast}\to
J/\psi+c\bar{c})}{\Gamma(\Upsilon\to\mathrm{light\; hadron})}=
\frac{3.85\alpha^2\alpha_s^2\langle\mathcal{O}^{\psi}_1(^3S_1)\rangle}{{6N_c(\frac{20}{243}\alpha_s^3(\pi^2-9)
+\sum_{q}2\pi e_q^2 e_b^2\alpha^2)m_b^3}}.
\end{equation}
By choosing the same numerical values for $r$, $m_b$, $e_q$, $\alpha$
$\langle\mathcal{O}^{\psi}_1(^3S_1)\rangle$ and
$\mathcal{B}(\Upsilon\to\mathrm{light\; hadron})$  as those in
Sec.III, the numerical result is
\begin{equation}
\mathcal{B}(\Upsilon\to\gamma^{\ast}\to
J/\psi+c\bar{c})=1.06\times10^{-6},
\end{equation}
and the normalized $J/\psi$ momentum spectrum is shown
in Fig.~3.

\subsection{$\Upsilon\to J/\psi+gg$}
The process $\Upsilon(p_{_0})\to J/\psi(p_{_1})+g(p_{_2})g(p_{_3})$
includes two parts, $\Upsilon\to\gamma^{\ast}\to
J/\psi+gg$ and $\Upsilon\to gg\gamma^{\ast}$ and $\gamma^{\ast}\to J/\psi$.
There are six Feynman diagrams for each part at the leading order with
the typical ones shown in Fig.~2b and 2c. To calculate the contribution of
the two parts together, the differential decay width is represented as
\begin{eqnarray}
&&\frac{d\Gamma}{d|\vec{p}_{_1}|}(\Upsilon\to J/\psi+gg)=
\frac{ 32\pi C_A C_F e_c^2e_b^2\alpha^2\alpha_s^2
\langle\Upsilon|\mathcal{O}_1(^3S_1)|\Upsilon\rangle \langle\mathcal{O}^{\psi}_1(^3S_1)\rangle
\sqrt{x_1^2-4r^2}}
{9(2N_c)^2 m_b^6\,r^3\,x_1\,\left( -1 + r^2 \right) \,{\left( 2\,r^2 - x_1 \right) }^3\,{\left( -2 + x_1 \right) }^3}
\nonumber\\&&
   ( ( -1 + r) \,( 1 + r ) \,( 2\,r^2 - x_1 ) \,( -2 + x_1 ) \,{\sqrt{-4\,r^2 +
{x_1}^2}}\,
       ( 8 + 8\,r^8 - 4\,r^6\,( -4 + 3\,x_1 )  +
       \nonumber\\&&
        r^4\,( -2 + x_1 ) \,( -16 + 7\,x_1 )  +
         x_1\,( -12 + ( 7 - 2\,x_1 ) \,x_1 )  -
         2\,r^2\,( -1 + x_1 ) \,( 8 + ( -7 + x_1 ) \,x_1 )  )  +
         \nonumber\\&&
      2\,( 1 + r^2 - x_1 ) \,( -( ( 2\,r^2 - x_1 ) \,
            ( 8 + 2\,r^8 + x_1\,( -12 + 5\,x_1 )  + r^6\,( 40 + x_1\,( -32 + 5\,x_1 )  )
            +\nonumber\\&&
              r^4\,( 6 - ( -2 + x_1 ) \,x_1\,( -19 + 6\,x_1 )  )  +
              r^2\,x_1\,( -6 + x_1\,( 13 + 2\,( -5 + x_1 ) \,x_1 )  )  )
              \nonumber\\&&
            \log (\frac{-2 + x_1 - {\sqrt{-4\,r^2 + {x_1}^2}}}{-2 + x_1 + {\sqrt{-4\,r^2 + {x_1}^2}}}) )  +
         r^2\,( -2 + x_1 ) \,( 8\,r^{10} - 12\,r^8\,x_1 +
            {x_1}^2\,( 5 + 2\,( -3 + x_1 ) \,x_1 )  +
            \nonumber\\&&
            r^6\,( 6 + x_1\,( -6 + 5\,x_1 )  )  + r^4\,( 40 + x_1\,( -38 + 13\,x_1 )  )  +
            r^2\,( 2 + x_1\,( -32 + ( 31 - 10\,x_1 ) \,x_1 )  )  )
            \nonumber\\&&
          \log (\frac{-2\,r^2 + x_1 + {\sqrt{-4\,r^2 + {x_1}^2}}}{-2\,r^2 + x_1 - {\sqrt{-4\,r^2 + {x_1}^2}}}) )
          ),
\label{eqn:jpsigg}
\end{eqnarray}
Where there is $x_1=\sqrt{|\vec{p}_{_1}|^2+4m_c^2}/m_b$.
And the normalized decay width becomes
\begin{equation}
\Gamma^{gg}_{\mathrm{Normal}}=\frac{\Gamma(\Upsilon\to
J/\psi+gg)}{\Gamma(\Upsilon\to\mathrm{light\; hadron})}=
\frac{60.8\alpha^2\alpha_s^2\langle\mathcal{O}^{\psi}_1(^3S_1)\rangle}{{6N_c(\frac{20}{243}\alpha_s^3(\pi^2-9)+\sum_{q}2\pi
e_q^2 e_b^2\alpha^2)m_b^3}}.
\end{equation}
By using the same parameters as above. We obtain
\begin{equation}
\mathcal{B}(\Upsilon\to J/\psi+gg)=1.67\times10^{-5}
\end{equation}
and the normalized $J/\psi$ momentum spectrum is plotted in Fig.~3.
In the numerical result, about $85.2\%$ contribution comes from the $\Upsilon\to gg\gamma^{\ast}(J/\psi)$ part,
$18.2\%$ from the $\Upsilon\to\gamma^{\ast}\to J/\psi gg$ part
and $-3.4\%$ from the interference part.

\section{Summary And Discussion}

To sum up all the contributions of the color-singlet QED and QCD
processes considered above, the branching ratio of direct $J/\psi$
production in $\Upsilon$ decay is
\begin{equation}
\mathcal{B_\mathrm{Direct}}(\Upsilon\to J/\psi+X)=3.9\times10^{-5},
\end{equation}
and the corresponding normalized $J/\psi$ momentum distribution is given
by the solid line in Fig.~3. It can be seen in Fig.~3 that
the contribution of the QCD process is dominated in small $p_\psi$ region, while
the effect of the QED process $J/\psi+gg$ is more important in large $p_\psi$ region.
In Eq.~(\ref{eqn:jpsigg}) and the dot-dashed line in Fig.~3, the
logarithmic divergence at the kinematic end point is obvious shown for the QED process $J/\psi+gg$.
It was pointed out in Ref~\cite{Fleming:2003gt,Lin:2004eu,Ma:2008gq} that both the $\alpha_{s}$
and $v_{b}$ expansion failed near the kinematic end point region in the similar processes $e^{+}e^{-}\to J/\psi+X$
and $\Upsilon\to \gamma+X$ because of the large perturbative and non-perturbative corrections, and the
logarithmic divergent behavior can be soften by applying the resummation in the SCET.
Whatever it can improve the
$J/\psi$ momentum spectrum largely near the kinematic end point, but the corrections to the total decay width is small.
Therefore we omit the resummation effect here.

\begin{figure}
\begin{center}
\includegraphics[scale=1.0]{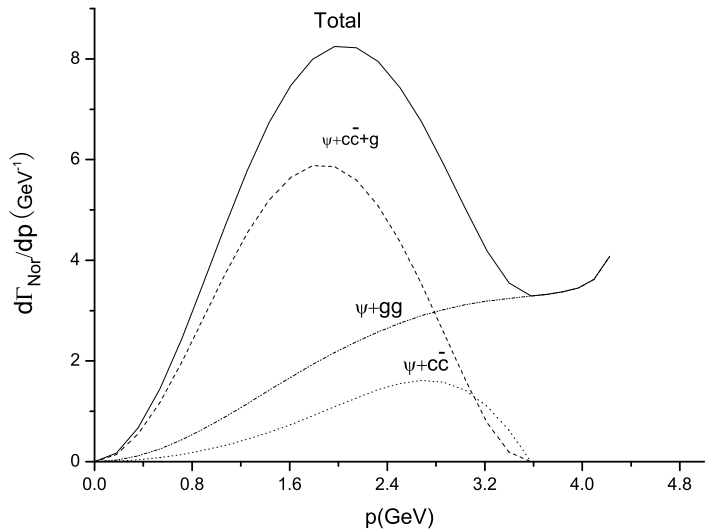}
\caption{ The contributions of QCD process $\Upsilon\to
J/\psi+c\bar{c}+g$(dashed line) and QED processes $\Upsilon\to
J/\psi+gg$ (dot-dashed line) and 5 times of
$\Upsilon\to\gamma^{\ast}\to J/\psi+c\bar{c}$ (dotted line) to $J/\psi$
momentum distribution for $J/\psi$ production in $\Upsilon$ decay.
And the sum of them is given by the solid line. }
\end{center}
\end{figure}

Our calculations show that at the leading order in $\alpha_{s}$, $v_{b}$
and $v_c$, the QCD process $\Upsilon\to J/\psi+c\bar{c}+g$ only
accounts for $54.4\%$ of the LO theoretical prediction for total
branching ratio, in spite of a enhancement factor
$\alpha_s^3/\alpha^2$ that is associated with the QCD and QED
coupling constants when compared to the QED processes. The main
reason lies on the fact that the virtuality of the two virtual
gluons are both of $m_b^{2}$ order in the QCD process while the
virtuality of the photon is fixed to $4m_c^2$ in the QED processes
dominated by $\Upsilon\to gg\gamma^{\ast}(J/\psi)$, and moreover the
four-body phase space of the QCD process is also less than the
three-body one of the QED processes.

On the experimental side, the CLEO collaboration find\cite{Briere:2004ug}
that the feed-down of $\chi_{cJ}$ to $J/\psi$ are $<8.2,11,10$ percent for $J=0,1,2$
respectively and the feed-down of $\psi(2S)$ is about $24$ percent in $\Upsilon \rightarrow J/\psi+X$.
Therefore it indicates that the experimental result of direct $J/\psi$ production would be
\begin{equation}
\mathcal{B_\mathrm{Direct}}(\Upsilon\to J/\psi+X)=3.52\times10^{-4}
\end{equation}
which is about 9 times larger than the presented theoretical results
based on the color-singlet calculations. This means that unlike the
conclusion before\cite{Li:1999ar} the branching ratio of
$\Upsilon\to J/\psi+X$ can not be explained by color-singlet model
at the leading order.

From the theoretical point of view, the color-octet mechanism can
account for most $J/\psi$ production, but its predictions for the
$J/\psi$ momentum spectrum is not agree with the experimental data.
The color-singlet predictions on the shape of the $J/\psi$ momentum
spectrum is more closer to the experimental result, but the
discrepancy of the branching ratio between them is large. For all
the numerical results, we used the theoretically normalized decay
width to estimate the branching ratio. Alternatively, by using
$\langle\Upsilon|O_1(^3S_1)|\Upsilon\rangle=2.9\mathrm{GeV^3}$\cite{Cheung:1996mh}
to calculate the partial decay width and choosing the total decay
width of $\Upsilon$  $51.4$ keV from the experimental
measurement\cite{Amsler:2008zzb}, the branching ratio will be
enhanced by a factor of about 3, which still can not explain the
experimental results. Therefore, it means that the NLO QCD
correction is important, just like in the known cases, the NLO QCD
corrections for $J/\psi$ production in $e^{+}e^{-}$ annihilation
show that the $K$-factor are about $1.97$ and $1.2$ for
$e^{+}e^{-}\to\gamma^{\ast}\to J/\psi+c\bar{c}$ and
$e^{+}e^{-}\to\gamma^{\ast}\to J/\psi+gg$ processes respectively;
the NLO QCD correction in $J/\psi$ related $\Upsilon$ exclusive
decays are also found quite important~\cite{Hao:2006nf}. In
addition, the contribution of $\mathcal{O}(\alpha_s^6)$ processes
$b\bar{b}(^3S_1,1)\to c\bar{c}(^3S_1,1)+gg$ and
$b\bar{b}(^3S_1,1)\to c\bar{c}(^3S_1,1)+gggg$ to the branching ratio
has been estimated to be of $10^{-4}$ order\cite{Trottier:1993ze}.
So that the next important step is to give an explicit and complete
calculations of them, which will be very helpful to understand the
conflict between the theory and experiment. Furthermore, to obtain
the full QCD correction for the inclusive $J/\psi$ production in
$\Upsilon$ decay would be a very interesting and challenge work for
explaining the experimental data. But it will involve very
complicated work at the QCD NLO and is beyond the scope of this
work.

\section*{\large{Acknowledgement}}
We thank Dr. S. Y. Li for helpful discussions. This work was
supported by the National Natural Science Foundation of China
(No.~10775141) and Chinese Academy of Sciences under Project No.
KJCX3-SYW-N2. Zhiguo He is currently supported by the CPAN08-PD14
contract of the CSD2007-00042 Consolider-Ingenio 2010 program, and
by the FPA2007-66665-C02-01/ project (Spain).

\end{document}